# Accelerated Corrosion of High Entropy Alloys under Tensile Stress

*Aditya Ayyagari[a]\*, Riyadh Salloom[a], Harpreet Singh Arora[b], Sundeep Mukherjee[a]*

[a.] *Department of Materials Science and Engineering, University of North Texas, Denton, Texas 76203, USA*
[b] *Department of Mechanical Engineering, School of Engineering, Shiv Nadar University, Uttar Pradesh 201314, India*
*\*Corresponding Author*: aa0715@unt.edu

**Abstract**

High entropy alloys are finding significant scientific interest due to their exotic microstructures and exceptional properties resulting thereof. These alloys have excellent corrosion resistance and may find broad range of applications from bio-implants, aerospace components and nuclear industry. A critical performance metric that determines the application worthiness of the alloys is the resilience of stressed structural members in a corrosive environment. This study reports the results from a novel experimental setup to quantify the corrosion rate under uniaxial tensile stress in a single phase fcc $Al_{0.1}CoCrFeNi$ high entropy alloy rods. Under a uniform uniaxial applied stress of 600 MPa, the corrosion current density was observed to increase by three orders of magnitude and ~150 mV drop in corrosion potential. The mechanism of accelerated corrosion is identified as surface passivation layer breakdown, pit initiation on un-passivated surface and rapid pit-propagation along the loading direction.

# Accelerated Corrosion of High Entropy Alloys under Tensile Stress

*Aditya Ayyagari[a], Riyadh Salloom[a], Harpreet Singh Arora[b], Sundeep Mukherjee[a]*

[a.] *Department of Materials Science and Engineering, University of North Texas, Denton, Texas 76203, USA*

[b] *Department of Mechanical Engineering, School of Engineering, Shiv Nadar University, Uttar Pradesh 201314, India*

There is an exponential growth in the high entropy alloys' research in recent time due to the large scope of alloy development possibilities [1]. Among the various alloy groups, the $Al_xCoCrFeNi_y$ has been particulary interesting as this alloy is versatile in terms microstructural evolution and unique properties there of. Minor variations in Al content transforms the microstructure from a simple single phase fcc alloy (x=0.1) to complex phase mixtures (x=0.3, 0.5) to a fully bcc matrix (at x=1.0) [2–5]. The system is also unique in terms of phase-decomposition products as it can also form a two phase eutectic system when the composition is adjusted to $AlCoCrFeNi_{2.1}$ [6]. The simplest among the systems is the $Al_{0.1}CoCrFeNi$ alloy that form a single phase fcc structure with exceptional corrosion resistance [7], remarkable phase stability [8], high resistance to irradiation damage[9] and excellent weldability[10]. However, the performance of this alloy system when accompanied by an external applied stress-field is unknown which is key to detemine its full application worthiness.

Corrosion behavior is significantly influenced by presence of stress-field. Although stress corrosion cracking is used for understanding the corrosion behavior of components exposed to stress and corrosive environments, the challenge with such an approach is the inability to predict the dynamic corrosion rates at different stress levels. This information is important to understand the detrimental (accelarating) effects of applied stress on corrosion. Here, we demonstarte a novel stress accelarated electrochemical (SAE) measurement system to measure the change in corrosion potential, corrosion current density and corrosion mechanisms at specific applied stress levels. The proposed set-up provides an oppurtunity to understand the mechanism of accelerated corrosion as surface passivation layer breakdown, pit initiation and rapid pit-propagation along the loading direction. The set-up and methodology is generic and can be applied to wide-range of material systems for designing highly corrosion resistance alloys.

A schematic of the test set up is shown in Figure 1. Gamry Ref3000 potentiostat in a three electrode configutration was used for measuring the electrochemical response from the sample. A saturated calomel electrode was used as referene electrode, high density graphite as counter electrode and the sample was connected as the working electrode. Uniaxial tensile loading component of the SAE module consisted of ADMET eXpert 4000™ professional table-top micro tensile testing machine. The two systems were operated by independent standalone computers and respective softwares. The sample in this study (Al$_{0.1}$CoCrFeNi alloy) was cast, forged and rod-rolled to 99.7% strain to produce a 2 mm diameter cylinder. The sample was annealed at 1200°C for 20 h to produce a fully recrystalized workpiece. Electron backscattered diffraction (EBSD) of the microstructural-crystallographic characterisation is shown in Figure 2. Figure 2a shows the EBSD map along the wire drawing direction where as Figure 2b shows it normal to the direction of drawing. Despite large plastic deformation, no preferred wire-drawing texture was senn and the mircostructure of the rod was highly homogenous and charecterized by the equi-axed strain-free grains. The pole figures conforming uniform disrtribution of grains in the principle crystallographic axes are shown in Figure 2b.

Baseline tensile properties of the wiredrawn material were established by performing the tensile test in lab air with a relative humidity of 22%, 24.5 °C, at a strain rate of 0.1 mm per minute. The tensogram of the annealed rod material is shown in supplimental Figure 2c, and the metrics are annotated. The properties of the wire material were identical to bulk cast-wrought material found in the literature [11]    Following baseline measurements, stress accelarated corrosion experiments were carried out by mounting the SAE module onto the tensile testing machine. Fresh set of samples were placed in the fixtures, the electrolytic cell was filled with 3.5 wt% NaCl slolution and appropriate connections were made as shown in Figure 1. The sample was allowed to attain stable open circuit potential (OCP) and subsequently swept through a rate of 0.16 mV/s at OCP. The limiting potential for the scan was set as the pitting potential based on observations from previous iterations. The corrosion response of the HEA rod under no external applied stress was recorded as $E_{corr}^{\sigma=0} = -235 \pm 16 \ mV$ and the $i_{corr}^{\sigma=0} = 16 \pm 6.5 \ nA/\text{cm}^2$ over six iterations. Fresh set of rod samples and electrolyte were used for each experimental iteration. Subsequently, the electrochemical signature while the sample was subjected to uniaxial tensile loading was recorded. Figure 3a shows the stress strain curve of the rod with the SAE module enveloped around it. The dynamic changes occuring in the OCP as the rod deformed are shown in Figure 3b. The first potential drop of 30 mV was seen at $E_{OCP}^{\sigma:242} = 299.8\ mV$ which corresponded to the on set of plastic deformation. The small potential drop was followed by recovery in OCP as the stress continued to increase up to 700 MPa where a sharp drop of 225 mV ($E_{OCP}^{\sigma:700}$ = -500 mV) was observed. This was immediately followed by a second potential recovery regime during which the OCP remained stable at -400 mV. The sample was stressed up to the ultimate tensile limit of 755 MPa at which third significant potential drop of 37 mV was recorded in the potentiogram. The OCP at UTS was $E_{OCP}^{\sigma:755}$ = -392 mV with a cummulative drop of 125 mV as compared to the pristine rod. The tensile loading was paused at UTS and the sample was subjected to a potential sweep to record the corrosion current density and corrosion potential. The corrosion response is shown in Figure 3c. The $i_{corr}^{\sigma:755} = 4.7 \pm 0.65\ \mu A/cm^2$ with a the slope of the anodic brach of the potentiogram slightly tending towards active corrosion. The electrochemcial test was terminated at current density of ~10 mA/cm$^2$.

Figure 4a shows the surface of the rod after corrosion experiments recorded in FEI-Quanta environmental scanning electron microscope. The sample was observed to have multiple cracks normal to the loading direction as shown in the magnified inset. Figure 4b shows surface cracks and consequent exposure of underlying base metal, akin to chareteristic mud-craking apperance. Elemental scans performed using inbuilt energy dispersive x-ray spectrometer showed that the surface oxide had cracked, exposing the underlaying material to electrolyte and consequent accelarated corrosion during the a potential sweep in Figure 3c. No distinct necking was noticed as the rod. The potential drops observed during OCP scan may have their origins in the surface oxide cracking. Sudden supply of electrons from the pristine material exposed due to surface oxide cracking may have resulted in drop in E$_{corr}$. This magnitude of potential drop was small at yield stress (~ 4% strain) where deformation was low, consequently having lower number of active sites that initiated corrosion. At flow-stress, (defined as a stress higher than the yielding point, but lower than fracture) the elongation was ~16% that likely resulted in exposure of a large fraction of upassivated surfaces which explains the large drop in the E$_{corr}$. This observation was consistent with multiple iterations of the experiment indicating that there may be a critical stress value between yield and tensile strengths at which rate of nucleation of deformation induced surface cracking is highest, and consequent accelerated corrosion. The recovery in OCP towards nobler value after sharp drops at yield, flow and ultimate tensile stresses may be due to the attainment of equilibrium with the electrolyte. Although there is recovery in the potential, there are still several active sites and larger surface area which might explain the relatively lower potetial recovery namely, higher active corrosion at higer stress levels.

High magnification SEM images of the corroded regions are shown in Figure 4c. Surface-oxide breakage and corrosion penetration from the unpassivated surface is shown in Figure 4d. It is observed that the kinetics of corrosion propagation is higher at the surface cracks tips, owing to the high stress concentration. The propagating crack tips coalese and form corrosion pits as shown in Figure 5c. A high magnification image of a large pit (Figure 4d) is shown in Figure 4e. Corrosion appears to propagate via a stepped morphology and preferentially along the loading direction. The direction of stress application of stress and resulting slip lines are clearly visible in the zoomed-in image Figure 4f. These slip planes and sharp boundaries indicate that preferred sites for corrosion nucleate along deforation bands akin to twins and grain boundaries in metalligraphic etching.

In conventional potentiodynamic polarization tests, where electrochemical parameters are obtained with no stress appplied to. While in stress corrosion cracking the electrochemical signals not are recorded. The experimental a module allowed the electrochemical parameters for an alloy under static or dynamic loading. This may be useful in determining the application worthiness of new structural materials such as HEAs subjected to stress under corrosive atmospheres.

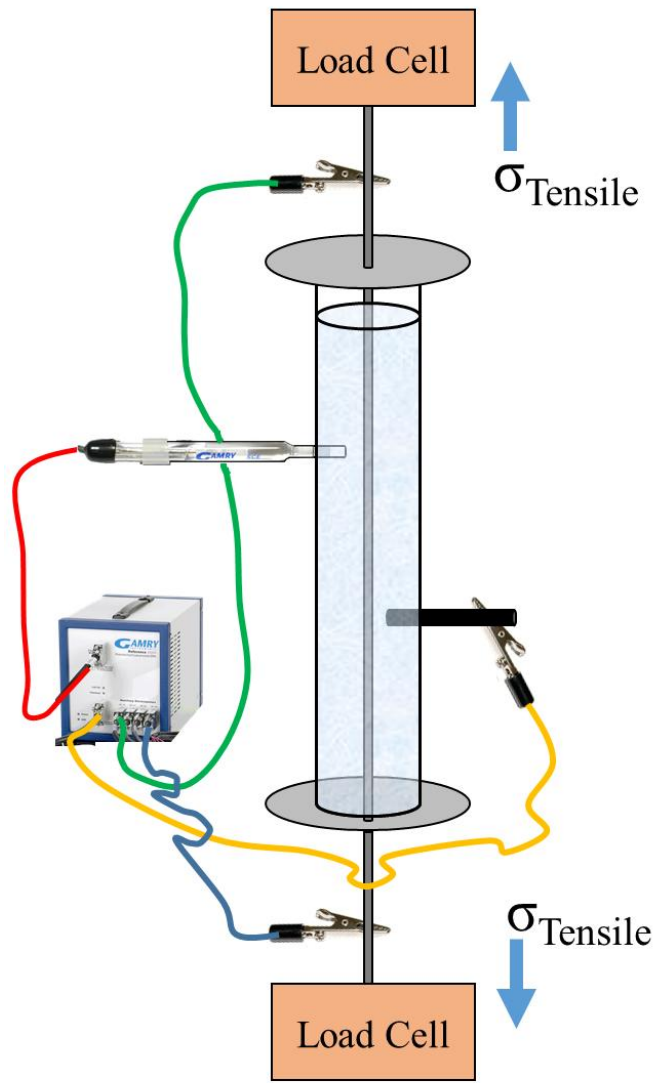

Figure 1: Schematic of SAE module showing HEA cylinder in the center. The rod is connected to Gamry potentiostat. The sample is the working electrode, graphite counter electrode and saturated calomel is the reference electrode.

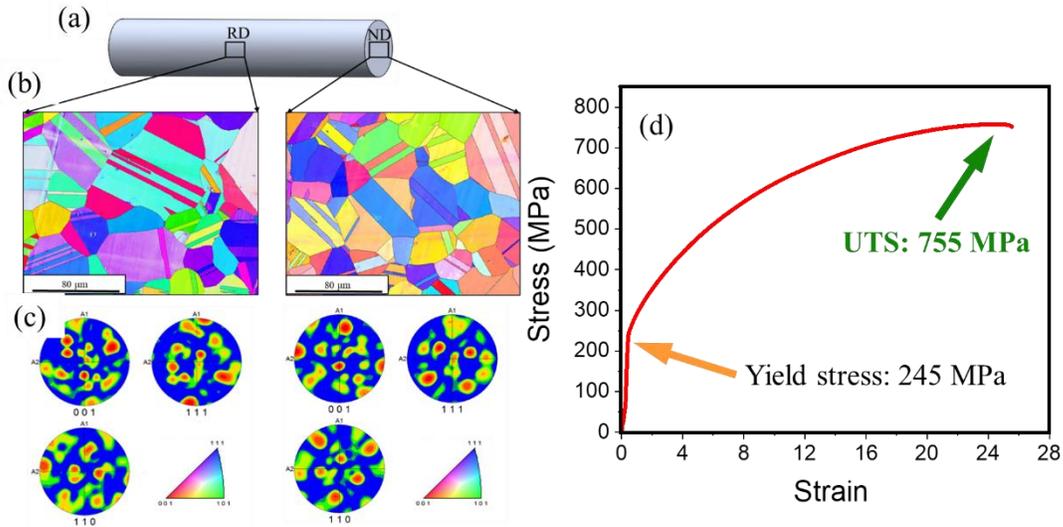

Figure 2: (a) Schematic of the metallic rod showing the rolling and normal directions (b) EBSD images of the HEA cylinder after cold rolling, rod drawing and static recrystallization annealing treatment. (c) pole figures showing distribution of grains in the principle crystallographic axes to be uniform (d) stress strain curve of showing the tensile properties of the samples.

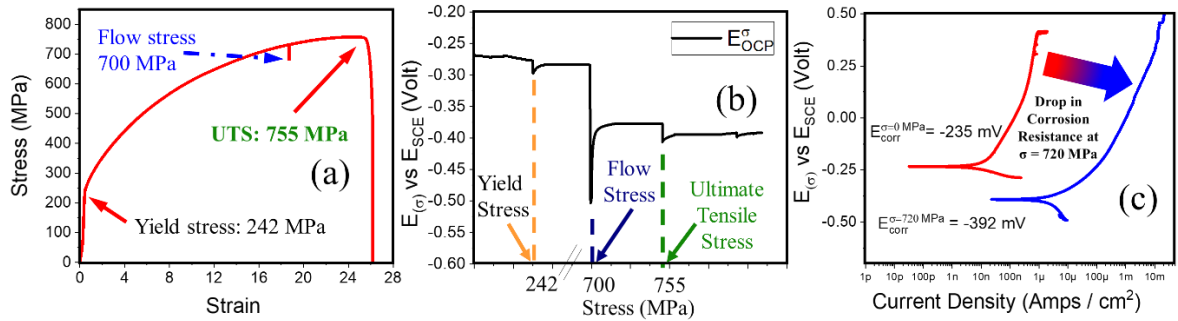

Figure 3: (a) Stress strain curve of the rod with the SAE module enveloped around it. (b) dynamic changes occuring in the OCP as the rod deformed, consisting of transitions between active and passive domains of corrosion (c) potentiodynamic polarization curves showing accerated corrosion behavior with increased stress

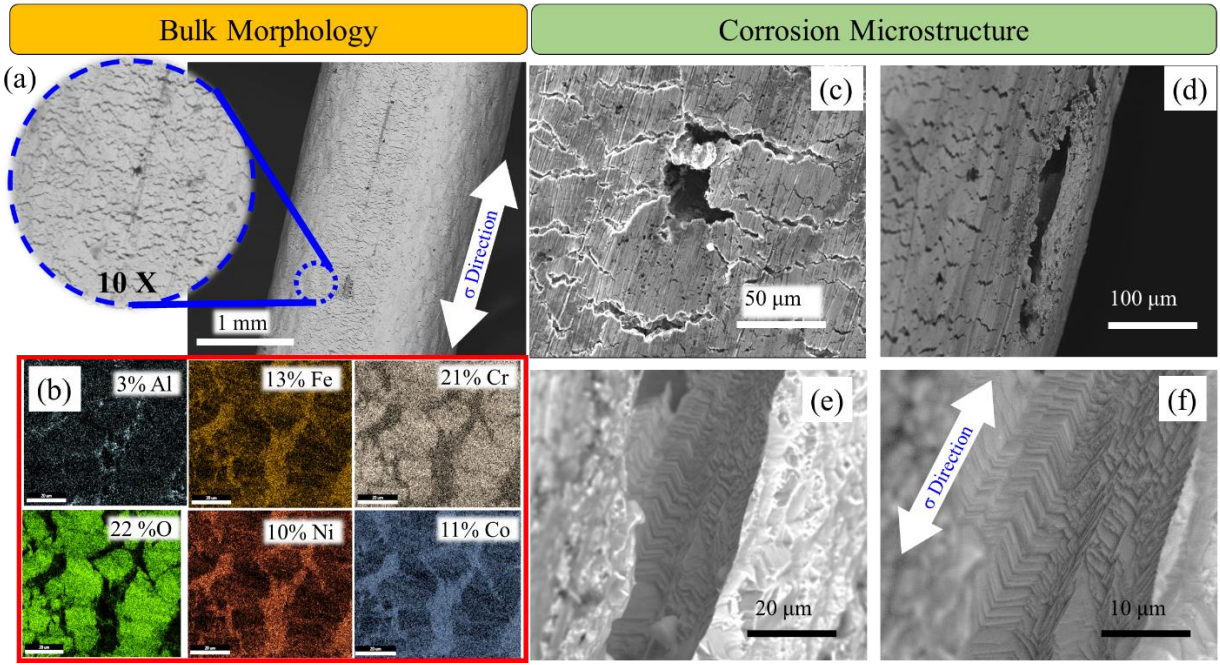

Figure 4: (a) Corrosion morphologies on the HEA surface showing oxide cracking; (b) EDS images show discontinuity of oxide regions and exposure of underlying metallic substrate (c) corrosion pits growing due to the coalescence of surface cracks (d) large surface crack (e) morphology of cracks propagating in the loading direction along the edges of slip planes (f) high magnification image of corroded slip planes.